\documentclass{emulateapj}
\usepackage{graphicx}
\usepackage{amsmath}
\usepackage{verbatim}
\usepackage{longtable}
\shorttitle{Sources in both H-ATLAS and {\it WISE}}
\shortauthors{Bond et al.}

\begin{document}

\title{THE INFRARED PROPERTIES OF SOURCES MATCHED IN THE {\it WISE} ALL-SKY AND HERSCHEL ATLAS SURVEYS}
\author{Nicholas A. Bond, Dominic J. Benford, Jonathan
  P. Gardner\altaffilmark{1}, Alexandre Amblard\altaffilmark{2}, 
  Simone Fleuren\altaffilmark{3}, Andrew W. Blain\altaffilmark{4}, Loretta Dunne\altaffilmark{5},  Daniel
  J. B. Smith\altaffilmark{6},  Steve J. Maddox\altaffilmark{5},
  Carlos Hoyos\altaffilmark{5}, Maarten Baes\altaffilmark{7}, David Bonfield\altaffilmark{6}, Nathan
  Bourne\altaffilmark{5},  Carrie Bridge\altaffilmark{8}, Sara Buttiglione\altaffilmark{9}, Antonio
  Cava\altaffilmark{10}, David Clements\altaffilmark{11}, Asantha Cooray\altaffilmark{12}, Ali Dariush\altaffilmark{13}, Gianfranco de Zotti\altaffilmark{14}, Simon
  Driver\altaffilmark{15,16}, Simon
  Dye\altaffilmark{5}, Steve Eales\altaffilmark{17}, Peter
  Eisenhardt\altaffilmark{18}, Rosalind Hopwood\altaffilmark{13}, Edo
  Ibar\altaffilmark{19}, Rob J. Ivison\altaffilmark{19}, Matt J. Jarvis\altaffilmark{6,20}, Lee
  Kelvin\altaffilmark{15,16}, Aaron
  S. G. Robotham\altaffilmark{15,16}, Pasquale Temi\altaffilmark{2}, Mark Thompson\altaffilmark{6},
  Chao-Wei Tsai\altaffilmark{21}, Paul van der Werf\altaffilmark{22}, Edward
  L. Wright\altaffilmark{23}, Jingwen Wu\altaffilmark{16}, Lin Yan\altaffilmark{24}}
\altaffiltext{1}{Cosmology Laboratory (Code 665), NASA Goddard Space Flight Center, Greenbelt, MD 20771}
\altaffiltext{2}{Astrophysics Branch, NASA/Ames Research Center, MS 245-6, Moffett Field, CA 94035}
\altaffiltext{3}{School of Mathematical Sciences, Queen Mary, University of London, Mile End Road, London, E1 4NS, UK}
\altaffiltext{4}{Department of Physics \& Astronomy, University of
  Leicester, University Road, Leicester LE1 7RH, UK} 
\altaffiltext{5}{School of Physics and Astronomy, University of Nottingham, University Park, Nottingham, NG7 2RD, UK}
\altaffiltext{6}{Centre for Astrophysics Research, Science \& Technology Research Institute, University of Hertfordshire, Hatfield, Herts, AL10 9AB, UK}
\altaffiltext{7}{Sterrenkundig Observatorium, Universiteit Gent, Krijgslaan 281 S9, B-9000 Gent, Belgium}
\altaffiltext{8}{Division of Physics, Mathematics, and Astronomy, California Institute of Technology, Pasadena, CA 91125}
\altaffiltext{9}{INAF-Osservatorio Astronomico di Padova, Vicolo Osservatorio 5, I-35122 Padova, Italy}
\altaffiltext{10}{Departamento de Astrof\'{\i}sica, Facultad de CC. F\'{\i}sicas, Universidad Complutense de Madrid, E-28040 Madrid, Spain}
\altaffiltext{11}{Imperial College, Astrophysics Group, Blackett Lab, Prince Consort Road, London, SW7 2AZ, UK}
\altaffiltext{12}{Department of Physics \& Astronomy, University of California, Irvine, CA 92697}
\altaffiltext{13}{Physics Department, Imperial College London, South Kensington Campus, SW7 2AZ, UK}
\altaffiltext{14}{INAF-Osservatorio Astronomico di Padova, Vicolo Osservatorio 5, I-35122 Padova, Italy, and SISSA, Via Bonomea 265, I-34136 Trieste, Italy}
\altaffiltext{15}{International Centre for Radio Astronomy Research (ICRAR), University of Western Australia, Crawley, WA 6009, Australia}
\altaffiltext{16}{SUPA, School of Physics and Astronomy, University of St. Andrews, North Haugh, St. Andrews, KY169SS, UK}
\altaffiltext{17}{School of Physics and Astronomy, Cardiff University,
  The Parade, Cardiff CF24 3AA, UK}
\altaffiltext{18}{Jet Propulsion Laboratory, California Institute of
  Technology, Pasadena, CA 91109} 
\altaffiltext{19}{UK Astronomy Technology Centre, Royal Observatory,
  Blackford Hill, Edinburgh EH9 3HJ, UK}
\altaffiltext{20}{Physics Department, University of the Western Cape, Cape Town, 7535, South Africa}
\altaffiltext{21}{IPAC, California Institute of Technology, Pasadena, CA 91125}
\altaffiltext{22}{Leiden Observatory, Leiden University, P.O. Box 9513, 2300 RA Leiden, The Netherlands}
\altaffiltext{23}{UCLA Astronomy, P.O. Box 951547, Los Angeles, CA
  90095-1547}
\altaffiltext{24}{Spitzer Science Center, California Institute of Technology, 1200 E. California Blvd., Pasadena CA 91125}

\begin{abstract}

  We describe the infrared properties of sources detected over
  $\sim\,$36~deg$^2$ of sky in the GAMA 15-hr equatorial field, using
  data from both the {\it Herschel} Astrophysical Terahertz Large-Area
  Survey (H-ATLAS) and Wide-field Infrared Survey ({\it WISE}).  With
  $5\sigma$ point-source depths of 34 and 0.048~mJy at $250\,\mu$m
  and $3.4\,\mu$m, respectively, we are able to identify $50.6$\% of
  the H-ATLAS sources in the {\it WISE} survey, corresponding to a
  surface density of $\sim 630$ deg$^{-2}$.  Approximately two-thirds
  of these sources have measured spectroscopic or optical/near-IR
  photometric redshifts of $z<1$.  For sources with spectroscopic
  redshifts at $z<0.3$, we find a linear correlation between the
  infrared luminosity at $3.4$~$\mu$m and that at $250$~$\mu$m, with
  $\pm 50$\% scatter over $\sim 1.5$ orders of magnitude in luminosity,
  $\sim$$10^9 - 10^{10.5}$~L$_{\odot}$. By contrast, the matched sources
  without previously measured redshifts ($r\gtrsim20.5$) have
  $250$-$350\,\mu$m flux density ratios that suggest either
  high-redshift galaxies ($z\gtrsim1.5$) or optically faint
  low-redshift galaxies with unusually low temperatures
  ($T\lesssim20$).  Their small $3.4$-$250\,\mu$m
  flux ratios favor a high-redshift galaxy population, as only the
  most actively star-forming galaxies at low redshift (e.g., Arp 220)
  exhibit comparable flux density ratios.  Furthermore, we find a
  relatively large AGN fraction ($\sim\,$30\%) in a $12\,\mu$m
  flux-limited subsample of H-ATLAS sources, also consistent with
  there being a significant population of high-redshift sources in the no-redshift
  sample.

\end{abstract}


\vspace{0.4in}

\section{INTRODUCTION}

The advent of submillimeter astronomy has opened a new window into the
Universe, allowing us to probe dusty star-forming galaxies at
high redshift, as well as cold dust in nearby galaxies.  Along with
the recent success of the balloon-borne BLAST experiment
\citep{BLAST,Eales09}, the launch of the {\it Herschel Space
  Observatory}\footnote[1]{{\it Herschel} is an ESA space observatory with
  science instruments provided by European-led Principal Investigator
  consortia and with important participation from NASA.}
\citep{Herschel} allows us to probe
wavelengths from $55$ to $672\,\mu$m to a spatial resolution of
$\lesssim\,$10\arcsec.  Its largest open-time key project, the {\it
  Herschel} Astrophysical Terahertz Large-Area Survey
\citep[H-ATLAS,][]{HATLAS}, will observe $\sim\,$550~deg$^2$
of sky and detect more than 300,000 galaxies.

Based upon a preliminary cross-identification with the Galaxy and Mass
Assembly survey \citep[GAMA,][]{GAMA} and Sloan Digital Sky Survey
\citep[SDSS,][]{DR7}, $\sim\,$50\%\ of the sources detected in the
H-ATLAS survey are at $z<1$ \citep{CrossIDs}.  In
addition, an analysis of the FIR colors of the remaining H-ATLAS
sources with flux densities above 35~mJy at $350\,\mu$m and $>3\sigma$
detections at $250\,\mu$m and $500\,\mu$m suggests an average redshift
of $z \sim2$ \citep{Amblard10,Lapi11}.  

Another method of constraining the redshift distribution of FIR-selected
sources is to look for counterparts in the near and mid infrared
(NIR and MIR).  Fleuren et al. (in prep) have performed source matching to
survey data from the VISTA Kilo-degree Infrared Galaxy survey
(VIKING, Sutherland et al., in prep), but these data only extend to $\sim\,$$
2\,\mu$m.  Starting in December 2009, the Wide-field Infrared
Survey Explorer\footnote[2]{For a description of the {\it WISE} mission, see
  http://wise2.ipac.caltech.edu/docs/release/prelim/expsup/}
\citep[{\it WISE},][]{WISE} began its mission to observe the entire sky in
four bands, ranging from $3.4$ to $22\,\mu$m, at $\lesssim\,$12\arcsec\
resolution.  Following the first public data release in April 2011,
$\sim\,$24,000~deg$^2$ of IR images and source catalogs became
available to the public, including $36$~deg$^2$ of sky in the
equatorial plane covered by the H-ATLAS survey.

The purpose of this letter is to describe the infrared properties of
identified {\it WISE} counterparts to H-ATLAS sources within the GAMA
15-hr field (G15).  Throughout we will assume a concordance cosmology
with $H_0=71$~km~s$^{-1}$~Mpc$^{-1}$, $\Omega_{\rm m}=0.27$, and
$\Omega_{\Lambda}=0.73$ \citep{WMAP}.  In a subsequent paper, we will
report WISE cross-identifications over the entire H-ATLAS area and
perform spectral energy distribution (SED) fits to the matched
sources.

\section{DATA AND METHODOLOGY}
\label{sec:data}

{\it Herschel} observations of the G15 field include imaging data at
$250$, $350$, and $500\,\mu$m from the SPIRE instrument, with
respective beam FWHM of $18\farcs1$, $25\farcs2$, and $36\farcs6$ \citep{SPIRE},
as well as $100$ and $160\,\mu$m imaging from the PACS instrument
\citep{PACS}.  The field subtends approximately 12$^{\circ}$ in right
ascension and 3$^{\circ}$ in declination and is centered on the
equatorial plane at $\alpha=14\,$h$\,30\,$m.  The H-ATLAS G15 source
catalogue (Dunne et al., in prep) is constructed using the same method as
the Science Demonstration Phase catalogue \citep{ATLAScatalog}.  It contains 27,481
sources detected at $>5\sigma$ in any of the three SPIRE bands and reaches
point-source depths of $34$, $40$, and $44$~mJy at $250$, $350$, and
$500\,\mu$m, respectively.

The {\it WISE} first public data release contains four-band coverage
of the entire G15 field to $5\sigma$ point-source depths of $0.048$,
$0.10$, $0.73$, and $5.9$~mJy at $3.4$, $4.6$, $12$, and $22\,\mu$m.
The angular resolution in these bands is $6\farcs1$, $6\farcs4$,
$6\farcs5$, and $12\farcs0$, respectively \citep{WISE}.  There are
$\sim\,$240,000 {\it WISE} sources within G15 detected at $>7\sigma$ in
at least one of the four {\it WISE} bands, corresponding to $\sim\,$0.14
{\it WISE} sources per $10$\arcsec-radius aperture.  In the {\it WISE}
preliminary release catalog, close pairs of sources are not deblended
for separations $\lesssim\,$9\arcsec, so when a background or foreground
source appears near an H-ATLAS source position, it will often be
blended with the true counterpart.

In addition to the infrared data from {\it WISE} and H-ATLAS, the G15
region has spectroscopic redshifts from the GAMA survey
\citep[$r\lesssim19.4$,][]{GAMA}, and photometric redshifts obtained
using optical/NIR photometry from SDSS, VIKING, and the UK Infrared
Deep Sky Survey Large Area Survey \citep[UKIDSS-LAS,][]{UKIDSS}.
These photometric redshifts are derived following \citet{CrossIDs} and
have typical redshift uncertainties of $\sigma_z/z\sim0.15$.  Of
the 27,481 H-ATLAS sources in the region, $16$\% have reliable
spectroscopic redshifts and $39$\% have photometric redshifts.

As a result of the high space density of $3.4\,\mu$m sources, a naive
matching to the H-ATLAS source positions will result in a
non-negligible number of misidentifications.  A significant fraction
of these misidentifications will be foreground stars that are blue in
all {\it WISE} bands and undetectable in H-ATLAS, so we can reduce the
contamination rate of our matched catalog by considering only {\it WISE}
sources with $[3.4]-[4.6]>0$.  The magnitude distribution of all
background/foreground {\it WISE} sources is such that $\sim\,$20\% are bluer
than this limit, as compared to $<1$\% of sources within $10$\arcsec\
of an H-ATLAS position.

To estimate the fraction of H-ATLAS sources with detectable
counterparts in the remainder of the {\it WISE} source catalog, we use the
method of Fleuren et al., in prep, where 
the detection rate is given by,
\begin{equation}
Q_0=1-\frac{\bar{S}}{\bar{B}},
\label{eq:Q0}
\end{equation}
where $\bar{S}$ is the fraction of unmatched positions in the H-ATLAS
catalog and $\bar{B}$ is the fraction of unmatched random positions.
Matching all {\it WISE} sources within $10$\arcsec\ (within which we
expect $>99$\% of the true matches to lie) and using $10^5$ random
positions within the G15 field, we find $Q_0=0.632 \pm 0.004$.  This
is larger than the value found in the SDSS galaxy catalog
\citep[$Q_0=0.583$,][]{CrossIDs}, but smaller than in the VIKING $Ks$
band ($Q_0=0.75$, Fleuren et al., in prep).  By contrast, we find
$Q_0=0.012 \pm 0.002$ within the subset of {\it WISE} sources with
$[3.4]-[4.6]<0$, suggesting that this color cut was effective in
removing stars.

\begin{figure}[t]
\plotone{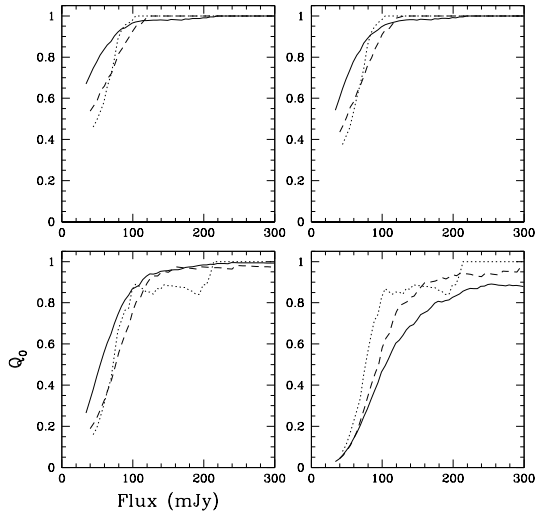}
\caption{NIR and MIR detection rates plotted as a function of limiting FIR flux density
  at $250\,\mu$m (solid), $350\,\mu$m (dashed), and $500\,\mu$m
  (dotted).  Detection rates are shown for $3.4\,\mu$m (upper left
  panel), $4.6\,\mu$m (upper right), $12\,\mu$m (lower left), and
 $22\,\mu$m (lower right) sources, where the corresponding $5\sigma$ 
detection limits are $0.06$, $0.9$, and $3.6$~mJy, respectively.  Note that the large sky density of
$3.4\,\mu$m sources can lead to ambiguity in the identification of some sources, so
identification rates are typically $\sim\,$10-20\% lower than
the corresponding detection rates.  
\label{fig:Completeness_v_Flux}}
\end{figure}

\begin{figure}[t]
\plotone{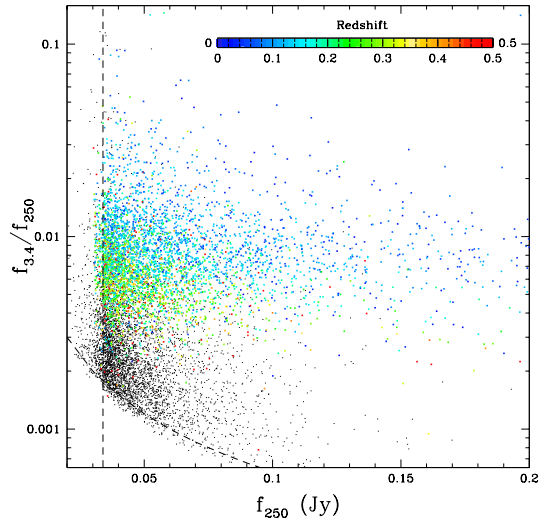}
\caption{Distribution of the $3.4$--$250\,\mu$m flux density ratio as a function
  of $250\,\mu$m flux density for H-ATLAS sources with {\it WISE} $3.4\,\mu$m IDs.
  Colored points have spectroscopic
  redshifts from the GAMA survey and sources without
  measured redshifts are indicated by black points.  Dashed lines indicate the
  approximate limits of the H-ATLAS and {\it WISE} surveys.  The
  majority of sources with spectroscopic redshifts ($r\lesssim19.4$) are well detected
  by WISE, while the no-redshift sources have small flux density
  ratios that are inconsistent with ``normal''
  low-redshift galaxies.\label{fig:FratvF}}
\end{figure}

In Fig.~\ref{fig:Completeness_v_Flux}, we use Eq.~\ref{eq:Q0} to
estimate the IR detection rate as a function of FIR flux density.  For
sources brighter than $100$~mJy at $250\,\mu$m, we find a {\it WISE}
$3.4\,\mu$m counterpart $>95$\% of the time.  The majority of
these sources ($78$\%) have measured spectroscopic or photometric
redshifts $<0.5$, suggesting that the low-redshift mode of
star-forming galaxies dominates sub-mm sources above this flux
density.
Below this FIR flux density, the NIR and MIR detection rates drop
rapidly as the high-redshift mode accounts for an increasing fraction
of the sub-mm sources (see Fig.~\ref{fig:FratvF}).

Approximately $1$\% of the objects in our $250\,\mu$m-selected sample
\citep{Negrello07}, including $\sim 50$\% of sources brighter than
$100$~mJy at $500\,\mu$m \citep{Negrello10,Hopwood11}, are expected to
be strongly-lensed $z>1$ galaxies.  Of the 46 such sources in the G15
region, $\sim\,$20\% are undetected at $12$ and $22\,\mu$m, most
likely because the $500\,\mu$m bandpass shifts higher in the blackbody
curve at high redshift, while the rest-frame MIR flux density declines
blueward of $22\,\mu$m \citep[e.g.,][]{Rieke09}.

The identification of individual NIR and MIR counterparts is more
subtle, as we want to eliminate as many of the false matches as
possible.  Here, we use the likelihood ratio technique of \citet{SS92},
which was implemented for the H-ATLAS survey in \citet[][hereafter
S11]{CrossIDs}.  
The likelihood that a given {\it WISE} source is
a counterpart to an H-ATLAS source is a function of the radial
probability distribution, $f(r)$, the $3.4\,\mu$m magnitude distribution of
non-counterparts, $n(m)$, and the $3.4\,\mu$m magnitude distribution of
detected H-ATLAS counterparts, $q(m)$:
\begin{equation}
L=\frac{q(m)f(r)}{n(m)},
\label{eq:L}
\end{equation}
where $f(r)$ is given by
\begin{equation}
f(r)=\frac{1}{2\pi \sigma_{\mathrm{
    pos}}^2}\mathrm{exp}\left(-\frac{r^2}{2\sigma_{\mathrm {pos}}^2}\right).
\label{eq:fr}
\end{equation}
We measure $n(m)$ directly from the {\it WISE} preliminary release
catalog, while we determine $q(m)$ by first measuring the magnitude distribution of $3.4\,\mu$m
sources within $10$\arcsec\ of the H-ATLAS source positions and then
subtracting the magnitude distribution of background/foreground sources normalized
by the area within this radius.  

Following S11, we assume the astrometric uncertainty to be circularly
symmetric and to have a dependence on the $250\,\mu$m signal-to-noise
ratio (SNR) given by,
\begin{equation}
\sigma_{\mathrm{pos}}=0.655\frac{\mathrm{FWHM}}{\mathrm{SNR}},
\label{eq:sig}
\end{equation}
where FWHM$=18.1$\arcsec\ is the {\it Herschel} beam width.  While the
{\it WISE} random astrometric uncertainties are negligible 
($\sim\,$$0\farcs3$), the {\it WISE} positions in the preliminary release
catalog are known to suffer from highly non-Gaussian systematic
offsets that can be as large as $1$\arcsec, so we apply a lower limit
of $\sigma_{\mathrm{pos}} >1.5$\arcsec.

Once we have computed the likelihood, $L_j$, for a possible counterpart $j$, we
determine the reliability by summing over all possible counterparts,
\begin{equation}
R_j=\frac{L_j}{\sum_iL_i+(1-Q_0)}.
\label{eq:R}
\end{equation}
We consider any $3.4\,\mu$m source with $R>0.8$ to be a reliable
counterpart.  After performing this procedure separately for {\it
  WISE} sources with star- and galaxy-like $[3.4]-[4.6]$ colors, we
obtain a combined total of 13,898 {\it WISE} counterparts, or $50.6
\pm 0.4$\% of the H-ATLAS catalog.  The expected number of false
matches can be obtained with
\begin{equation}
N_{\mathrm{falseID}}=\sum_i(1-R_i).
\label{equation}
\end{equation}
We estimate that there are 369 false identifications in our sample,
corresponding to a contamination rate of $2.7$\%.  The majority of our IDs have
galaxy-like colors, but we were able to identify $14$ objects with star-like colors
($[3.4]-[4.6]<0$) that weren't already identified as galaxies with
SDSS matching.  This corresponds to an upper limit of $0.05$\% for the
fraction of H-ATLAS sources identified as stars and detected in {\it WISE}.

Comparing the individual {\it WISE} identifications to SDSS
cross-identifications, we find that $84.7$\% of the $10,709$ G15 H-ATLAS
sources with $R>0.8$ SDSS identifications also had $R>0.8$
{\it WISE} identifications.  The median separation between the {\it WISE} and SDSS
sources is $0\farcs6$, consistent with the astrometric uncertainties
in the current version of the {\it WISE} catalog.  In addition, $\sim\,$6\% of
these sourcess have separations $>\,$3\arcsec, consistent with the
expected contamination rates of the {\it WISE} ($\sim\,$3\%) and SDSS ($\sim\,$5\%) matched samples.

\section{DEMOGRAPHICS}
\label{sec:demographics}

\begin{figure}[t]
\plotone{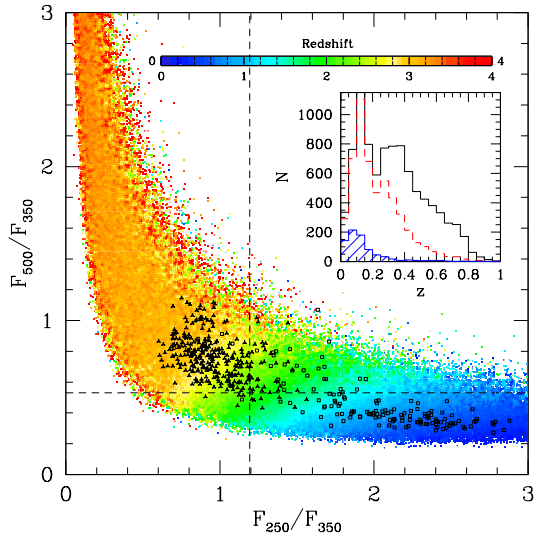}
\caption{FIR flux density ratio for H-ATLAS sources with {\it
    WISE} identifications.  Colored
  grid cells indicate the median redshift of a set of modified blackbody models with
temperatures uniformly distributed between 10 and 60~K and emissivity
parameters between 0 and 2.  Black points indicate sources with
$5\sigma$ detections in all three SPIRE bands, where open squares have spectroscopic
redshifts, $z<0.2$, and solid triangles have no spectroscopic or
photometric redshift measurement.  Dashed
lines indicate the median flux density ratios of a stack of sources lacking
redshifts.  The majority of the no-redshift sources appear to be
at $z\sim2$. {\it Inset}: Combined distribution of spectroscopic and
photometric redshifts for H-ATLAS sources with detections at $3.4\,\mu$m (solid histogram), $12\,\mu$m (dashed), and
$22\,\mu$m (hatched).  There are 73 sources at $z>1$ not
shown.  
\label{fig:FluxRats}}
\end{figure}

The redshift distribution of $250\,\mu$m-selected sources is likely
bimodal, with populations of both moderately-star-forming galaxies at
$z<1$ and high-redshift starburst galaxies at $z\sim 2$
\citep{CrossIDs,Lapi11}.  Because the latter are typically faint in
the observed-frame optical and UV, the vast majority of objects
with measured spectroscopic or photometric redshifts are in the
low-redshift mode.  Of the H-ATLAS sources with spectroscopic and
photometric redshifts (see the inset of Fig.~\ref{fig:FluxRats}), we
successfully identify $85.6$\% in {\it WISE}.

Although most of the {\it WISE} identifications with known redshifts
are at $z<0.8$ ($>\,$98\%), such a sample is biased toward low
redshift because redshift measurements require detectable flux in the
UV, optical, or near-infrared.  Approximately one-third ($30.4$\%) of
the identified sources have reliable spectroscopic redshift and
two-thirds ($66.5$\%) have photometric redshifts from optical/NIR
data.  For the remaining sources, we can estimate redshifts using
their FIR flux density ratios \citep[e.g.,][]{Amblard10,Shulz10},
which give an indication of the location of the peak of the FIR dust
emission.  In Fig.~\ref{fig:FluxRats}, we compare the SPIRE colors of
our matched catalog to a suite of $10^6$ modified blackbody spectral
energy distributions
\citep{Amblard10}.  We indicate on the plot the median flux density
ratios of the $4,677$ H-ATLAS sources with {\it WISE} identifications
but no redshifts (``no-redshift'' sample, dashed lines), as well as
the individual flux density ratios of 311 such sources with $>\,$3$\sigma$
detections in all three SPIRE bands.

The no-redshift sample has median FIR colors suggesting $z\sim2$,
while the subset with three-band SPIRE detections may have even higher
redshifts, perhaps as large as $z\sim3.5$.  The redshifts derived
using the \citet{Amblard10} technique are highly uncertain, as
redshift is degenerate with dust temperature and it is
possible that these galaxies have typical dust temperatures different
than $35$~K \citep[which is typical of high-z SMGs,
e.g.][]{Chapman05}.  However, in order for the no-redshift sample to
be at $z\sim0$, they would need to have very cold dust temperatures
($T\sim12$~K) in addition to being very faint in the optical
($r\gtrsim20.5$).  Although temperatures as low as $10$~K have been
seen in H-ATLAS galaxies at $z<1$, $T\sim25-30$~K is more typical
\citep{Dye10}.  If the no-redshift sample has median temperatures
typical of these low-redshift H-ATLAS galaxies, then our redshift
estimate drops to $z\sim 1.5$.

Further evidence that the no-redshift sources are at $z\gg0$ can be
found in the $3.4$-$250\,\mu$m ratios (see Fig.~\ref{fig:FratvF}).
H-ATLAS sources with GAMA redshifts have NIR-FIR flux density ratios
that decrease from $\sim\,$0.009 at $z\sim0$ to $\sim\,$0.006 at
$z\sim0.35$.  This trend, which continues toward higher redshift, is
due primarily to the larger $k$-correction at $3.5\,\mu$m than at
$250\,\mu$m \citep[e.g.,][]{Rieke09}.  The median flux density ratio
for the no-redshift sources, by comparison, is $0.0023$.  Although
low-redshift galaxies will occasionally exhibit such small NIR-FIR
flux density ratios -- Arp 220, for example, has
$f_{3.4}/f_{250}=0.0019$ -- such objects are rare, actively
star-forming, and unlikely to have dust temperatures as small as
$\sim\,$15~K.

\begin{figure}[t]
\plotone{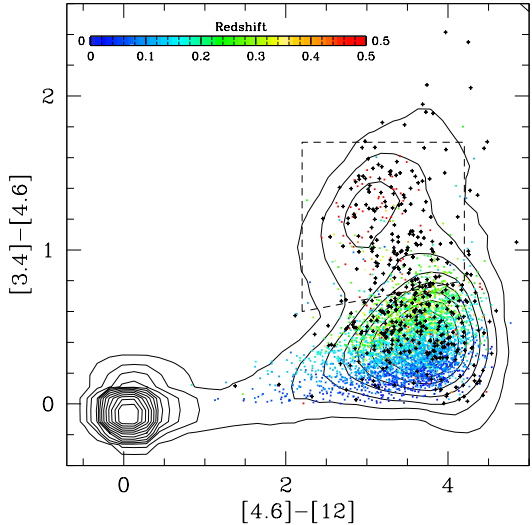}
\caption{IR color distribution of ATLAS-{\it WISE} matches with
three-band {\it WISE} photometry, including sources with $z<0.5$
(color coded by redshift) and no-redshift sources (black points).  The diagram can be divided into stars
(lower left corner), star-forming galaxies (lower
right corner), and AGNs (dashed region).  The color distribution of all
three-band {\it WISE} detections (including sources unmatched to H-ATLAS) is
indicated by contours.  Sources without measured redshifts tend to be redder in
$[3.4]-[4.6]$ and have a larger AGN fraction than sources with
spectroscopic or photometric redshift measurements, $z<0.5$.
\label{fig:WISEcolors}}
\end{figure}

We can shed even further light on the properties of the no-redshift
sample by examining their position in {\it WISE} color space (see
Fig.~\ref{fig:WISEcolors}).  The technique is similar to those
developed for a set of IRAC filters at comparable wavelengths
\citep{SLS05,Stern05}.  It was demonstrated that the diagram can be
divided into stars and early-type galaxies (lower left corner,
$[4.6]-[12]\lesssim1$), star-forming galaxies (lower right corner),
and active galactic nuclei (AGNs), where AGNs are selected largely
based on their power-law emission in the NIR and MIR.  This was
adapted for {\it WISE} by \citet{Jarrett11},
\begin{align}
[4.6]-[12]>2.2 \nonumber \\
[4.6]-[12]<4.2 \nonumber \\
[3.4]-[4.6]<1.7 \label{eq:AGNcrit}\\
[3.4]-[4.6]>0.1([4.6]-[12])+0.38\nonumber.
\end{align}
Of the 4,959 H-ATLAS sources with {\it WISE} identifications, $z<0.5$, and
detections in $3.4$, $4.6$, and $12\,\mu$m, the majority are
star-forming galaxies, with $<0.2$\% main sequence stars \citep[which is consistent
with the SDSS/H-ATLAS matching done by][]{Thompson10}, and a small
AGN fraction ($0.057$),

For the $9$\% of no-redshift sources with three-band {\it WISE} photometry,
we find a much larger AGN fraction ($0.30$), but are subject to a
selection bias toward AGNs due to their additional flux from warm dust
at $12\,\mu$m \citep{Hainline11}.  Furthermore, emission from
polycyclic aromatic hydrocarbons in high-redshift star-forming galaxies can
occasionally mimic AGNs in their IRAC colors, leading to contamination
in our AGN samples \citep[e.g.][]{Lacy04,Donley12}.
For comparison, \citet{Coppin10} performed SED fits on a set of blank-field SMG
samples and estimated that $\sim\,$15\% of SMGs are dominated by an AGN
in the mid-infrared. 

\section{FIR-NIR PROPERTIES}
\label{sec:colors}

Emission from dust in the FIR is often used as an indicator of the
total star formation rate in galaxies \citep[e.g.,][]{Kennicutt98},
under the assumption that most of the dust heating is provided by
young stars and that the star-forming regions are optically thick.
The $250\,\mu$m bandpass samples the cooler dust emission, which can
arise from both star-forming clouds and diffuse regions in the ISM
\citep[e.g.,][]{Eales10,Dunne11}.  By contrast, $3.4\,\mu$m emission is
dominated by stellar sources and is a probe of the stellar mass for galaxies
without a strong AGN contribution \citep[e.g.,][]{Stern05}.  

\begin{figure}[t]
\plotone{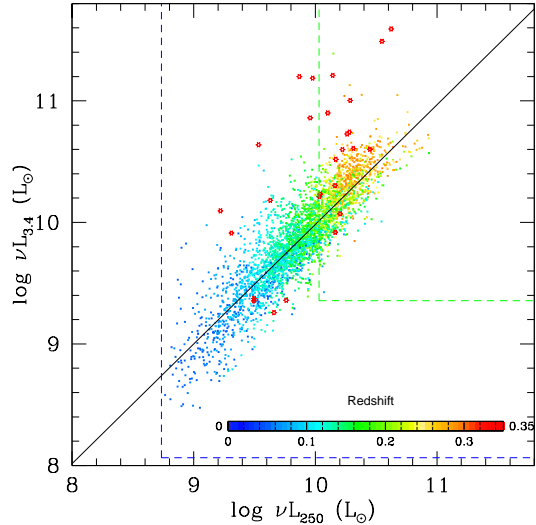}
\caption{Luminosity at $3.4\,\mu$m as a function of luminosity at
  $250\,\mu$m for H-ATLAS sources with spectroscopic redshifts,
  $0.05<z<0.3$.  Both luminosities are rest-frame $k$-corrected.  
Small squares are colored according to their
  redshift and red stars indicate AGN, selected with the criteria
  given in Eq.~\ref{eq:AGNcrit}.  Dashed lines indicate the
  approximate selection limits for sources at $z=0.05$ and $0.2$.  The majority of the AGN
lie above the relationship, suggesting that $L_{3.4}$ contains a
non-negligible contribution from hot dust emitting near the central
black hole.\label{fig:LvsL}}
\end{figure}

In Fig.~\ref{fig:LvsL}, we plot $3.4\,\mu$m luminosity as a function
of $250\,\mu$m luminosity for H-ATLAS sources with spectroscopic
redshifts, $0.05<z<0.3$.  We $k$-correct $3.4\,\mu$m using a power law
interpolation of the $[3.4]-[4.6]$ color.  For the $250\,\mu$m flux
densities, the majority of the galaxies in our sample do not have
sufficient FIR flux density to obtain reliable dust temperatures, as
is needed for a proper $k$ correction.  Instead, we use the median
dust temperature of $26$~K found by \citet{Dye10} using PACS and SPIRE
data from the H-ATLAS SDP of 1,346 $0.1<z<0.5$ sources.  Although the
$r<19.4$ limit of the GAMA spectroscopic survey restricts the range of
luminosities that can appear on the plot, we estimate that GAMA misses
only $\sim 5$\% of matched $z<0.3$ galaxies based upon the
deeper ($r\lesssim20.5$) subsample of H-ATLAS galaxies with
photometric redshifts.

We find that the majority of the IR-color-selected AGN (red points)
lie above the relationship for star-forming galaxies, as $L_{3.4}$
will contain a non-negligible contribution from hot dust emitting near
the central black hole in addition to emission from low-mass stars.
However, if we exclude AGNs, we find an approximately linear
correlation, with a best-fit power law index,
$\alpha=0.98^{+0.03}_{-0.05}$.  The quoted systematic uncertainties on
the power law index were determined by allowing for a range of
possible dust temperatures, $18<T<34$~K, when performing $250\,\mu$m
$k$-corrections.  The intrinsic scatter about this relationship is
$0.18 \pm 0.01$~dex, or $\sim\,$50\%.

The existence of a linear correlation between $L_{250}$ and $L_{3.4}$
across one and a half decades in luminosity suggests a close relationship between
the cold dust probed by the FIR and the stellar mass probed by the
NIR.  Previous indications with {\it Herschel} have shown the
$250\,\mu$m luminosity density to tightly correlate with both the
$24\,\mu$m luminosity density, a star formation rate indicator, and
the total infrared luminosity \citep{Elbaz10}.  This fact, coupled
with a linear correlation between stellar mass and star formation rate
for the general star-forming galaxy population \citep[][Donoso et al.,
in prep]{Daddi07,Elbaz07,Noeske07}, suggests that the majority of the
low-redshift H-ATLAS sources are actually ``normal'' star-forming
galaxies.

\acknowledgments

This publication makes use of data products from the Wide-field
Infrared Survey Explorer, which is a joint project of the University
of California, Los Angeles, and the Jet Propulsion
Laboratory/California Institute of Technology, funded by the National
Aeronautics and Space Administration.

{\it Herschel} is an ESA space observatory with science instruments
provided by European-led Principal Investigator consortia with sig-
nificant participation from NASA. U.S. participants in {\it
  Herschel} ATLAS acknowledge support provided by NASA through a
contract issued from JPL.

In addition, we thank Dan Stern and Roberto Assef for helpful
discussions about the quasar selection.

\bibliographystyle{apj}                       

\bibliography{apj-jour,Bond0409}

\renewcommand{\thefootnote}{\alph{footnote}}

\end{document}